\documentclass[fleqn,usenatbib]{mnras}

\usepackage[T1]{fontenc}

\DeclareRobustCommand{\VAN}[3]{#2}
\let\VANthebibliography\thebibliography
\def\thebibliography{\DeclareRobustCommand{\VAN}[3]{##3}\VANthebibliography}

\usepackage{graphicx}	
\usepackage{amsmath}	

\usepackage{newtxtext,newtxmath}

\title[Chemical Kinetics with Neural Networks]{Using a neural network approach to accelerate disequilibrium chemistry calculations in exoplanet atmospheres}

\author[J.L.A.M. Hendrix et al.]{
Julius L.A.M. Hendrix,$^{1}$\thanks{E-mail: jlam.hendrix@gmail.com}
Amy J. Louca,$^{1}$
Yamila Miguel$^{2,1}$
\\
$^{1}$Leiden Observatory, Leiden University, Postbus 9513, 2300 RA Leiden, The Netherlands\\
$^{2}$SRON Netherlands Institute for Space Research, Niels Bohrweg 4, 2333 CA Leiden, the Netherlands
}

\date{Accepted XXX. Received YYY; in original form ZZZ}

\pubyear{2022}

\begin{document}
\label{firstpage}
\pagerange{\pageref{firstpage}--\pageref{lastpage}}
\maketitle

\begin{abstract}
In this era of exoplanet characterisation with JWST, the need for a fast implementation of classical forward models to understand the chemical and physical processes in exoplanet atmospheres is more important than ever. Notably, the time-dependent ordinary differential equations to be solved by chemical kinetics codes are very time-consuming to compute. In this study, we focus on the implementation of neural networks to replace mathematical frameworks in one-dimensional chemical kinetics codes. Using the gravity profile, temperature-pressure profiles, initial mixing ratios and stellar flux of a sample of hot-Jupiter atmospheres as free parameters, the neural network is built to predict the mixing ratio outputs in steady state. The architecture of the network is composed of individual autoencoders for each input variable to reduce the input dimensionality, which is then used as the input training data for an LSTM-like neural network. Results show that the autoencoders for the mixing ratios, stellar spectra, and pressure profiles are exceedingly successful in encoding and decoding the data. 
Our results show that in 90\% of the cases, the fully trained model is able to predict the evolved mixing ratios of the species in the hot-Jupiter atmosphere simulations. The fully trained model is $\sim 10^3$ times faster than the simulations done with the forward, chemical kinetics model while making accurate predictions. 

\end{abstract}




\section{Introduction}
There are two methods commonly used for calculating the abundance of different species in an atmosphere: thermochemical equilibrium and chemical kinetics 
(\citeauthor{bahn1960} \citeyear{bahn1960}; \citeauthor{zeleznik1968} \citeyear{zeleznik1968}). Thermochemical equilibrium calculations treat each species independently and do not require an extensive list of reactions between different species. Consequently, this method is fast for estimating the abundance of different species in an exoplanet atmosphere and has been widely used in the community (e.g., \citeauthor{Stock2018} \citeyear{Stock2018}; \citeauthor{Woitke2018} \citeyear{Woitke2018}). However, the atmospheres of exoplanets are dynamic environments. Both physical and chemical processes can alter the compositions and thermal structures of the atmosphere. In particular, atmospheric processes like photochemistry, mixing and condensation of different species can affect atmospheric abundances, deviating the concentrations observed from what would be found by chemical equilibrium calculations (\citeauthor{Cooper2006} \citeyear{Cooper2006}; \citeauthor{Swain2008} \citeyear{Swain2008}; \citeauthor{moses11} \citeyear{moses11}; \citeauthor{Kawashima2021} \citeyear{Kawashima2021}; \citeauthor{Roudier2021} \citeyear{Roudier2021}; \citeauthor{Baxter2021} \citeyear{Baxter2021}). For example, the recent detection of SO\textsubscript{2} (\citeauthor{Feinstein2022} \citeyear{Feinstein2022}; \citeauthor{Ahrer2022} \citeyear{Ahrer2022}; \citeauthor{Alderson2022} \citeyear{Alderson2022}; \citeauthor{Rustamkulov2022} \citeyear{Rustamkulov2022}) and the determination of this species as direct evidence of photo-chemical processes shaping the atmosphere of WASP 39b (\citeauthor{Tsai2022} \citeyear{Tsai2022}), suggest that certain exoplanet atmospheres are in disequilibrium and we need chemical disequilibrium models using chemical kinetics to correctly interpret the observations.

Chemical kinetics codes consider the effects that lead to a non-equilibrium state in the atmosphere. These codes incorporate a wide range of atmospheric processes 
such as the radiation from the host star that can drive the dissociation of molecules  --or photochemistry--, the mixing of species at different pressures due to the planet's winds, or the diffusion of species, and calculate the one-dimensional abundances of species in exoplanetary atmospheres (e.g. \citeauthor{moses11} \citeyear{moses11}; \citeauthor{Venot12} \citeyear{Venot12}; \citeauthor{Miguel2014} \citeyear{Miguel2014}; \citeauthor{Tsai17} \citeyear{Tsai17}; \citeauthor{Hobbs19} \citeyear{Hobbs19}). However, 
To calculate the abundance of different species using chemical kinetics, a system of coupled differential equations involving all the species must be solved, and prior knowledge of reaction rates and a reaction list is necessary to estimate the production and loss of each species. Therefore, as more species and reactions are incorporated into the chemical networks, the complexity of these simulations increases, and so does the computational cost these simulations require. The result is that chemical kinetics codes have long computational times, and can not be used by more detailed calculations (e.g. circulation models) or as a fast way of interpreting observations (by retrieval codes), which are usually subject to simplifications.

For the past few decades, the use of machine learning techniques, specifically neural networks (NN), has become more prevalent in research fields outside of computer science. Within astronomy, neural networks have been used for applications like image processing \citep{Dattilo19}, adaptive optics \citep{Landman21}, exoplanet detection \citep{Shallue2018}, exoplanetary atmospheric retrieval \citep{Cobb2019} and chemical modelling \citep{Holdship21}, and more traditional machine learning techniques have been used for applications like exoplanetary atmospheric retrieval \citep{nixon2020} and chemistry modelling of protoplanetary disks \citep{Smirnov-Pinchukov2022}. Trained neural networks are fast to use, so a neural network trained to accurately reproduce the outcomes of chemical kinetics codes could greatly reduce computational time. Such a neural network could simulate a large amount of atmospheric conditions in a short period of time, which is for example useful for atmospheric retrievals from observational constraints. It could also be incorporated into a multi-dimensional atmospheric simulation that connects a multitude of individual one-dimensional simulations by the implementation of atmospheric mixing and other global processes.

In this study, we investigate the feasibility of machine learning techniques for speeding up a one-dimensional chemical kinetics code. To this end, we perform calculations on a fiducial giant planet as an example to show how this technique can be used to bring the best of these two worlds: the detailed information of chemical kinetics calculations and the speed of Neural Networks techniques. In the next section, we explain in more detail how we obtain the dataset and the specifics of the architectures used. The results of our networks are presented in the following section (section \ref{sec:results}), which are discussed afterwards in section \ref{sec:discussion}. Finally, we summarise and conclude our findings in section \ref{sec:conclusion}.

\section{Methods}
\subsection{Chemical Kinetics}\label{sec:vulcan}
Chemical kinetics is the most realistic way of calculating abundances and is necessary, particularly at low temperatures (T < 2000 K) and pressures (P < 10 -- 100 bars), where timescales of processes such as mixing in the atmosphere are shorter than chemical equilibrium and dominate the chemistry and abundances in the atmosphere.

We make use of the one-dimensional chemical kinetics code VULCAN (\citeauthor{Tsai17} \citeyear{Tsai17}; \citeyear{Tsai21}), to create a large dataset on the atmospheres of gaseous exoplanets. The code is validated for hot-Jupiter atmospheres from 500 K to 2500 K. VULCAN calculates a set of mass differential equations:
\begin{equation}
\frac{\partial n_i}{\partial t}= \mathcal{P}_i - \mathcal{L}_i-\frac{\partial \Phi_i}{\partial z},
\end{equation}
where $n_i$ is the number density of the species $i$, $t$ is the time, $\mathcal{P}_i$ and $\mathcal{L}_i$ are the production and loss rates of the $i$-th species, and $\Phi_i$ its transport flux that includes the effects of dynamics caused by convection and turbulence in the atmosphere. For a more complete derivation of this equation from the general diffusion equation, we refer the reader to \cite{Hu2012}. VULCAN starts from initial atmospheric abundances calculated using the chemical equilibrium chemistry code \textit{FastChem} \citep{fastchem}, although we note that the final disequilibrium abundances are not affected by the choice of initial values adopted \cite{Tsai17}, and further evolves these abundances by solving a set of Eulerian continuity equations that includes various physical processes (e.g. vertical mixing and photochemistry). To solve these partial differential equations, VULCAN numerically transforms them into a set of stiff ordinary differential equations (ODEs). These ODEs are solved using the \textit{Rosenbrock} method, which is described in detail in the appendix of \cite{Tsai17}. In this study, we make use of machine learning techniques to solve these equations and hence speed up the process.

\subsection{Building the dataset}\label{sec:dataset}

\subsubsection{Parameter Space}

To construct the dataset, we vary the following parameters:

\begin{enumerate}
    \item \textbf{Planet mass, M} [M$_\mathrm{J}$]: within the range [0.5, 20] M$_\mathrm{J}$.
    \item \textbf{Orbit radius, r} [AU]: within the range [0.01, 0.5] AU.
    \item \textbf{Stellar radius, $\rm R_{\star}$} [R$_{\odot}$]: within the range [1, 1.5] R\textsubscript{$\odot$}.
\end{enumerate}

Other parameters such as surface gravity, irradiation temperature, and stellar effective temperature are derived from these free parameters.

\begin{enumerate}
    \item \textbf{Planet radius} [R\textsubscript{Jup}]: This is derived from the planet mass using the relation from \cite{Chen17}, shown in Equation \ref{eq:planet_mass_radius}, where $R$ is the planet radius and $M$ is the planet mass:
\end{enumerate}
\begin{equation}
    \frac{R}{R_{\oplus}} = 17.78 \left( \frac{M}{M_{\oplus}} \right)^{-0.044} .
    \label{eq:planet_mass_radius}
\end{equation}
We note that our aim is to present the results for a simple general case, and the mass-radius relation we use is suitable for this purpose. However, we must emphasize that the relation between mass and radius for giant exoplanets is not unique and depends on various factors, such as the mass of metals, core mass, irradiation received by the planet, and their effect on the inflation of radius. All of these factors can impact the evolution path of giant planets and their final radius, leading to a dispersion in the mass and radius relation.
\begin{enumerate}
    \item \textbf{Temperature-pressure profile}: As our aim is to demonstrate the use of neural networks for calculating non-equilibrium chemical abundances in a general case, we have utilized an analytical, non-inverted temperature-pressure profile from \cite{Heng14}. While these analytical profiles are simplistic, they are widely used in the literature to explore general cases and are suitable for our purposes. However, for calculating the chemistry of a real planet, more detailed calculations that take into account the opacities of different species and their abundances in the atmosphere should be included. The assumptions for this calculation are $T_{int} = 120$ K, $\kappa_L = 0.1$, $\kappa_S=0.02$, $\beta_S = 1$ and $\beta_L =  1$, based on the default values included in the VULCAN code \cite{Tsai17}. The pressure profile is constructed within the range [$10^{-2}$, $10^9$] dyne cm$^{-2}$. This calculation is an important step, as it determines whether the set of parameters is valid for the dataset. If any part of the temperature profile falls outside of the range [500, 2500] K, the temperature range for which VULCAN is validated, the example is rejected from the dataset.    
    
    \item \textbf{Stellar flux}: the stellar spectra used for the dataset have two sources: the Measurements of the Ultraviolet Spectra Characteristics of Low-mass Exoplanetary Systems (MUSCLES) collaboration (\citeauthor{muscles1} \citeyear{muscles1}; \citeauthor{muscles2} \citeyear{muscles2}; \citeauthor{muscles3} \citeyear{muscles3}) and the PHOENIX Stellar and Planetary Atmosphere Code \citep{phoenix}. The MUSCLES database contains observations from M- and K-dwarf exoplanet host stars in the optical, UV, and X-ray regime, and is used for stars with an effective temperature lower than 6000 K. For effective temperatures of 6000 K and above, stellar spectra are generated by the PHOENIX model. Flux values below $10^{-14}$ erg nm$^{-1}$ cm$^{-2}$ s$^{-1}$ are cut-off.
\end{enumerate}

The remaining parameters of the VULCAN configuration files are kept constant throughout the dataset. Eddy- and molecular diffusion are both taken into account as fixed parameters. For the eddy diffusion constant, we make use of a constant of $K_{zz} = 10^{10}$ $\mathrm{cm^2/s}$. The molecular diffusion constant is taken for a hydrogen-dominated gas as described in \cite{banks}. As standard chemical network, we make use of VULCAN's reduced default N-C-H-O network that includes photochemistry. We assume solar elemental abundances for the hot-Jupiters, and we make use of 150 pressure levels for the height layers. The output of VULCAN is saved every 10 steps. In total, 13291 valid configurations are generated within the parameter space. 

\begin{figure*}
    \centering
    \includegraphics[scale=0.18]{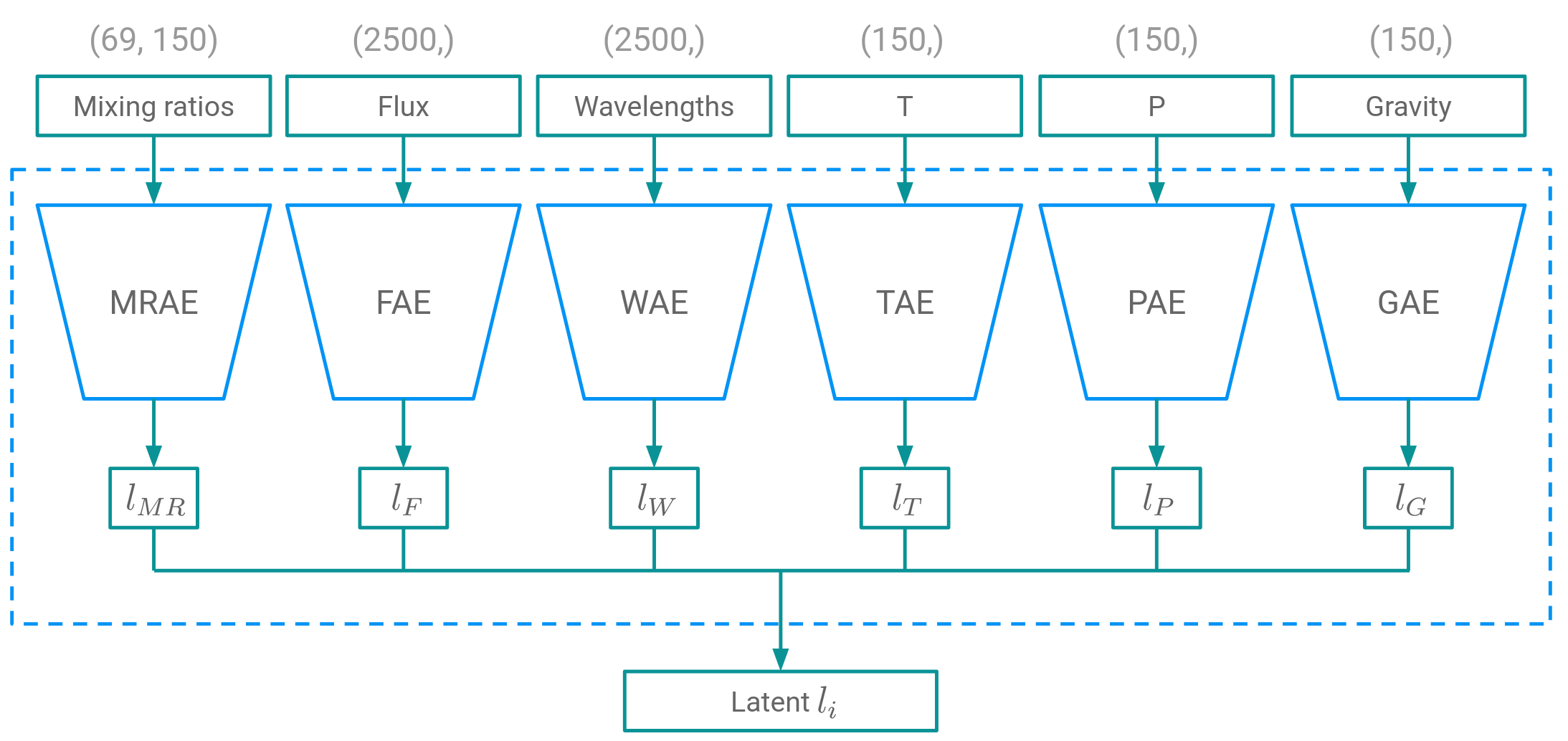}
    \caption{The different properties and their corresponding encoders that together encode the input data. The decoding process is symmetrical to this encoding process, where every property has a corresponding property decoder. }
    \label{fig:encoders_overview}
\end{figure*}

\subsubsection{Formatting}\label{subsec:formatting}

In order to limit the computation times when training the network,
the input-output pairs do not contain all of the information supplied by VULCAN. 

For the inputs, a selection of six properties is made. These properties are extracted from VULCAN before time integration starts, so they can be interpreted as the initial conditions of the simulation. The six properties are:

\begin{enumerate}
    \item \textbf{Initial mixing ratios}: the initial mixing ratios of the species in the simulation. These mixing ratios are calculated by VULCAN using FastChem \citep{fastchem}. The shape of the array containing the mixing ratios is (69, 150), as the mixing ratios are defined for 69 species for 150 height layers each.
    
    \item \textbf{Temperature profile}: the temperature profile as calculated by the analytical expression from \cite{Heng14}. The temperature is defined for every height layer, so it has a shape of (150,).
    
    \item \textbf{Pressure profile}: the pressure profile that is calculated as part of the temperature profile calculation. It is of the same shape, (150,).
    
    \item \textbf{Gravitational profile}: the gravitational acceleration per height layer. It has the shape (150,).
    
    \item \textbf{Stellar flux component}: one of the two components that make up the stellar spectrum contains the flux values. This is generated from either the MUSCLES database and/or the PHOENIX model and interpolated to a shape of (2500,).
    
    \item \textbf{Stellar wavelength component}: the second component of the stellar spectrum contains the wavelengths corresponding to the flux values. It has the same shape, (2500,).
\end{enumerate}

For the outputs, we make use of the time-dependent mixing ratios.

Because not every simulation takes the same amount of time to converge to a solution, the number of saved abundances differs per VULCAN simulations. To include the information contained in the evolution of the abundances through time, 10 sets of abundances, including the steady-state abundances, are saved in each output. This set of abundances is evenly spaced through time, so the simulation time between abundances will vary for different VULCAN simulation runs. Before the abundances are saved, they are converted to mixing ratios. The shape of the outputs is (10, 69, 150).

\subsubsection{Data Standardisation}

The inputs and outputs of the various components differ by several orders in magnitude. To ensure that the neural network trained on the data set is not biased towards higher-valued parameters, the data has to be standardised. First, the distributions of the properties are standardised according to Equation \ref{eq:dist_standardization}:

\begin{equation}
    p_{s} = \frac{\mathrm{log}_{10} (p) - \mu }{\sigma},
    \label{eq:dist_standardization}
\end{equation}

with

\begin{equation}
    \mu = \frac{1}{n} \sum_{i=0}^n \mathrm{log}_{10} (p_i),
\end{equation}

and

\begin{equation}
    \sigma = \sqrt{\frac{1}{n} \sum_{i=0}^n \left( \mathrm{log}_{10} (p_i) - \mu \right) ^2},
\end{equation}

where $p$ is the property to be scaled, $n$ is the size of the dataset and $p_s$ is the standardised property. After standardisation, the properties are normalised in the range [0, 1]:

\begin{equation}
    p_{s,n} = \frac{p_{s} - \mathrm{min}(p_{s})}{\mathrm{max}(p_{s}) - \mathrm{min}(p_{s})},
\end{equation}

where $p_{s,n}$ is the final normalised property.

Once the input properties are normalised, the output mixing ratios are normalised with the same scaling parameters as were used for the input mixing ratios. When the trained neural network is presented with an input for which to predict the mixing ratios, it only has information about the scaling parameters of the inputs. To be able to unnormalise the outputs, they need to be scaled with the same scaling parameters.

\begin{figure*}
    \centering
    \includegraphics[scale=0.2]{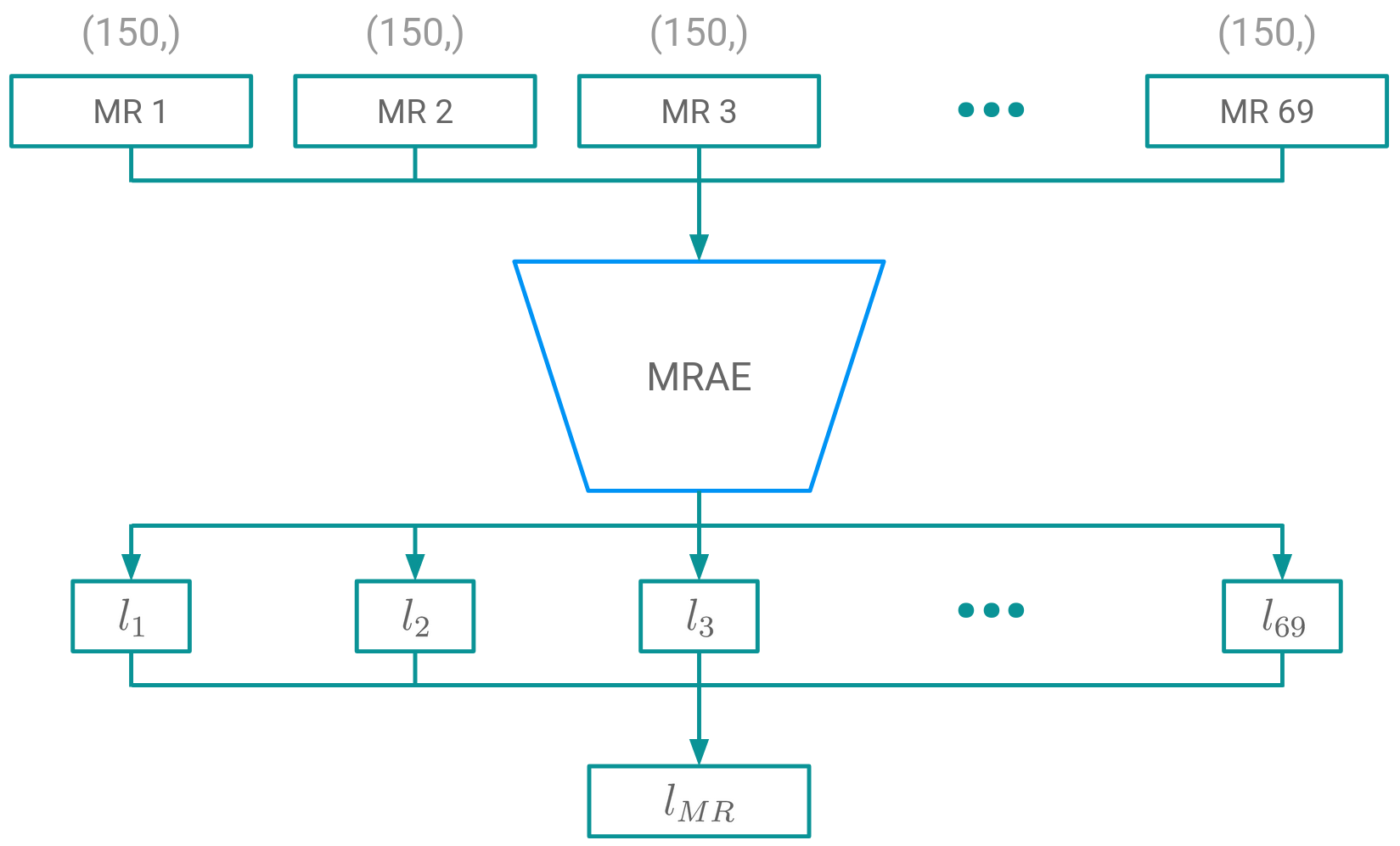}
    \caption{A more detailed sketch of the architecture of the mixing ratio autoencoder. MR $i$ denotes the mixing ratio of a certain species, $i$, for all height layers, and $l_i$ denotes the encoded mixing ratios for species $i$.}
    \label{fig:MRAE}
\end{figure*}

\subsection{Model Architecture}\label{sec:model_architecture}

\subsubsection{Autoencoder Structure}\label{subsec:AE}

The input of each configuration within the dataset consists of roughly 15800 values. To speed up the training process and complexity of the neural network we make use of an \textit{autoencoder} (AE) for reducing the dimensionality of the examples in the dataset. In previous studies this approach has been shown an effective way to reduce dimensionality within chemical kinetics (e.g. \citeauthor{Grassi2022} \citeyear{Grassi2022}).
An autoencoder consists of two collaborating neural networks: an \textit{encoder} and a \textit{decoder}. The task of the encoder is to reduce the dimensionality of the input data by extracting characterising features from the example and encoding them in a lower dimensionality representation called the \textit{latent representation}. The task of the decoder is to take the latent representation and use it to reconstruct the original input data with as little loss of information as possible. The encoder and decoder are trained simultaneously, and no restraints are placed on the way the autoencoder uses its \textit{latent space}, apart from the size of the latent representations.

As is discussed in Section \ref{subsec:formatting}, the inputs of the model consist of six properties. Because these properties do not share the same shape, we cannot encode and decode them using a single autoencoder. Instead, we construct six unique autoencoders, one for each property of the model inputs. Figure \ref{fig:encoders_overview} shows an overview of the process of encoding the initial conditions. The decoding process is not shown but is symmetrical to the encoding process.
Each encoder conceals a specific property into a corresponding latent representation. To get the latent presentation of the entire input example, $l_i$, the property latent representations $\{l_{MR}, l_F, l_W, l_T, l_P, l_G\}$ are concatenated. When decoding the latent representation of the input, the latent vector $l_i$ is split back into the different property latent representations, and given to that property's decoder. Every encoder-decoder pair is trained separately.

The hyperparameters of each autoencoder are optimized by trial and error. A summary of each set of hyperparameters is shown in table \ref{tab:ae_hyperparameters}.

\subsubsection*{Mixing Ratios Autoencoder}
The mixing ratios are the largest contributor to the size of the model inputs. Compressing each of the species' mixing ratios efficiently reduces the size of the input latent representation, $l_i$, by a substantial amount. Because of the limited size of the training dataset, a compromise has to be made to successfully train this autoencoder. Rather than concurrently encoding all 69 species for each example, each species is encoded individually. As a result, the training dataset expands by a factor of 69, while disregarding any potential correlations in species abundances during the encoding procedure. 

Figure \ref{fig:MRAE} shows the application of such an autoencoder: for a given input, each of the 69 species' mixing ratios is encoded into corresponding latent vectors $\{ l_1, l_2, l_3, ..., l_{69}\}$. The concatenation of these 69 latent vectors then makes up the latent representation of the mixing ratios $l_{MR}$.

All encoders and decoders are multilayer perceptron (MLP) neural networks. For the mixing ratio autoencoder (MRAE), the encoder and decoder both consist of 7 fully connected layers, followed by hyperbolic tangent activation functions. The encoder input layer has a size of 150 and the output layer has a size of 30. The hidden layers have a size of 256. Adversely, the decoder has an input layer size of 30 and an output layer size of 150. The compression factor of the MRAE is therefore $150/30=5$.

To train the MRAE, the dataset is split into a train dataset (70\%), a validation dataset (20\%), and a test dataset (10\%). 
To increase the size of the dataset, the MRAE is trained on a shuffled set\footnote{Using a random sampler function within the PyTorch package.} of the mixing ratios of both the inputs and the output of the chemical kinetics simulations. The performance of the autoencoder is measured using the loss function in equation \ref{eq:relative MSE}:

\begin{equation}
    \mathcal{L} = \frac{1}{N} \sum_{i=1}^{N} \left ( \frac{p_i - a_i}{a_i} \right)^{2}.
    \label{eq:relative MSE}
\end{equation}

where $\mathcal{L}$ is the loss, $N$ is the number of elements in the actual/predicted vector, and $p_i$ and $a_i$ are $i$-th elements of the predicted and actual vectors, respectively. The MRAE is optimised using the Adam optimiser \citep{adam}, with a learning rate of $10^{-5}$. A batch size of 32 is used, and the model is trained for 200 epochs. These hyperparameters can also be found in table \ref{tab:ae_hyperparameters}.

\begin{table*}
    \centering
    \caption{Hyperparameters for the property autoencoders. Each autoencoder has an encoder and decoder neural network. These are MLPs, consisting of 7 fully connected layer with hyperbolic tangent activation functions.}
    \begin{tabular}{c|c|c|c|c|c|c}
        \textbf{model} & \textbf{hidden size} & \textbf{latent size} & \textbf{optimiser} & \textbf{learning rate} & \textbf{batch size} & \textbf{epochs}\\
        \hline
        \hline
        MRAE & 256 & 30 & Adam & $10^{-5}$ & 32 & 200 \\
        PAE & 256 & 2 & Adam & $10^{-6}$ & 4 & 100 \\
        TAE & 256 & 30 & Adam & $10^{-5}$ & 4 & 100 \\
        GAE & 256 & 30 & Adam & $10^{-5}$ & 4 & 100 \\
        FAE & 1024 & 256 & Adam & $10^{-5}$ & 4 & 200 \\
        WAE & 1024 & 2 & Adam & $10^{-7}$ & 4 & 200 \\
    \end{tabular}
    \label{tab:ae_hyperparameters}
\end{table*}

\subsubsection*{Atmospheric profile Autoencoders}
The temperature, pressure, and gravity profiles all have the same shape of (150,), so their autoencoders can use the same architecture. Moreover, the atmospheric profiles share their shape with the mixing ratios of individual species (i.e. the height layers in the atmosphere). Therefore, a very similar neural network structure to that of the MRAE is used for the atmospheric profile autoencoders. The encoder input layer shape, the decoder output layer shape, and the hidden layer shapes are taken directly from the MRAE for all atmospheric profile autoencoders. An important parameter to tune for each atmospheric profile autoencoder separately is the size of the latent representations.

The pressure profile is set logarithmically 
for all examples in the dataset. By taking the logarithm of the pressures the spacing becomes linear, $\log(P_i) - \log(P_{i+1})$. Theoretically, we then only need two values to fully describe the pressure profile: the pressure at the first and last height layers. To encode these values, no autoencoder is needed. One could take this one step further, and provide input parameters like mass and radius, from which the pressure and gravity profile are dependent, directly to the core model as inputs. While this more specialised approach is suitable for these input parameters, it is not generalisable to other input parameters. To keep the model architecture more general and adaptable to different input parameters, an autoencoder is used nonetheless. The size of the pressure profile autoencoder (PAE) latent representations is set to 2. This corresponds to a compression factor of $150/2=75$.

The temperature and gravity profiles are not linear. For both the temperature autoencoder (TAE) and gravity autoencoder (GAE), a latent representation size of 30 is used. This corresponds to a compression factor of $150/30=5$, the same as for the MRAE.

All profile autoencoders are evaluated using the loss function previously defined in equation \ref{eq:relative MSE} and are optimised using the Adam optimiser. The TAE and GAE use a learning rate of $10^{-5}$, and the PAE uses a learning rate of $10^{-6}$. All profile autoencoders are trained with a batch size of 4, for 100 epochs (see also table \ref{tab:ae_hyperparameters}).

\subsubsection*{Stellar Spectrum Autoencoders}

After the mixing ratios, the stellar spectrum components contribute predominantly to the input data size.  
The stellar spectrum is comprised of a flux and a wavelength component. These components share the same shape, so one NN structure can be used for both autoencoders. 
The structure of the encoder and decoder is similar to that of the MRAE: a 7-layer, fully connected MLP, with hyperbolic tangent activation functions after each layer. The encoder input layer and decoder output layer have a size of 2500, and the hidden layers have a size of 1024.

Similarly to the PAE, the wavelength bins are spaced logarithmically. Again only two values are needed to fully describe the wavelength range. 
The latent representation size for the wavelength autoencoder (WAE) is, therefore, also 2. The compression factor for this network is $2500/2=1250$. The flux autoencoder (FAE) has a latent representation size of 256, which gives it a compression factor of $2500/256 \approx 10$.

Both autoencoders are evaluated using the loss function from equation \ref{eq:relative MSE}. They are optimised using the Adam optimiser, the WAE with a learning rate of $10^{-7}$, and the FAE with a learning rate of $10^{-5}$. They are both trained for 200 epochs with batches of 4 examples (see also table \ref{tab:ae_hyperparameters}).

\subsubsection{Core Network}\label{subsec:core_network}

\begin{figure}
    \centering
    \includegraphics[width=\linewidth]{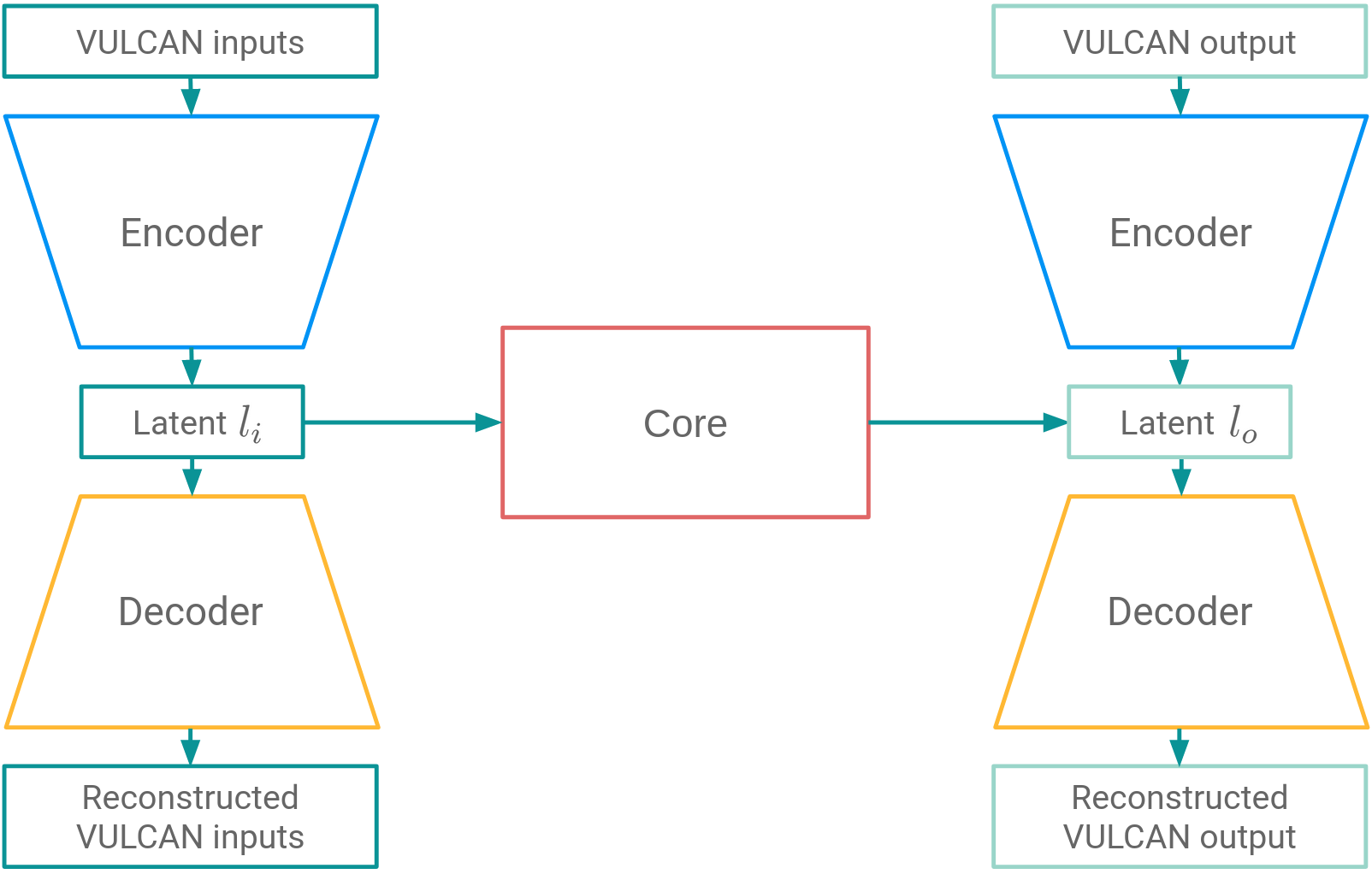}
    \caption{An overview of the model architecture. The core neural network maps between the latent representations of the VULCAN inputs $l_i$ and outputs $l_o$.}
    \label{fig:model_overview}
\end{figure}

As mentioned before, the outputs are also large in dimensionality. Because the outputs contain mixing ratios for all species, for 10-time steps (Section \ref{sec:dataset}), they can be encoded using the MRAE. 
Figure \ref{fig:model_overview} shows how the autoencoder would encode both the inputs and the last time step of the outputs to their latent representations $l_i$ and $l_o$, respectively. Note that even though the autoencoder is shown twice in this figure, the same autoencoder is used to encode both the inputs and the outputs. In the middle of the figure, connecting the two latent spaces, a second neural network called the \textit{core network} is located.

The function of the core network is to learn a mapping between the latent representations of the inputs and the evolved outputs. The design of the core network takes advantage of some of the characteristics of VULCAN. From Section
    \ref{sec:vulcan} we know that VULCAN solves ODEs for specific atmospheric configurations over a simulated period of time. To impart this sense of time in the core neural network, a \textit{Long-Short Term Memory} (LSTM) is used as the base of the design. The LSTM was chosen for its proven performance in numerous applications, from stellar variability (e.g. \citeauthor{Jamal_2020} \citeyear{Jamal_2020}) to Core-collapse supernovae search (e.g. \citeauthor{Alberto2023} \citeyear{Alberto2023}) and solar radio spectrum classification (e.g. \citeauthor{Xu2019} \citeyear{Xu2019}), as well as the ease of implementation. The LSTM has known shortcomings like the vanishing gradient problem and long training times when dealing with long sequences. However, with the short sequence length used with our model (i.e. 10 timesteps), these shortcomings are not considered problematic for this proof of concept.

The input of the core network is not sequential in nature. With some changes, we can use the LSTM in a 'one-to-many' configuration. In this configuration, the initial output of the LSTM $h_0$ is given to an MLP. This MLP produces a vector with the same shape as the initial input $x_0$, which can be interpreted as the 'evolved' input $x_1$. This evolved input is fed back into the LSTM to produce $h_1$, from which the MLP produces $x_2$, and so forth. This can be repeated for an arbitrary number of steps. 

\begin{figure*}
    \centering
    \includegraphics[scale=0.2]{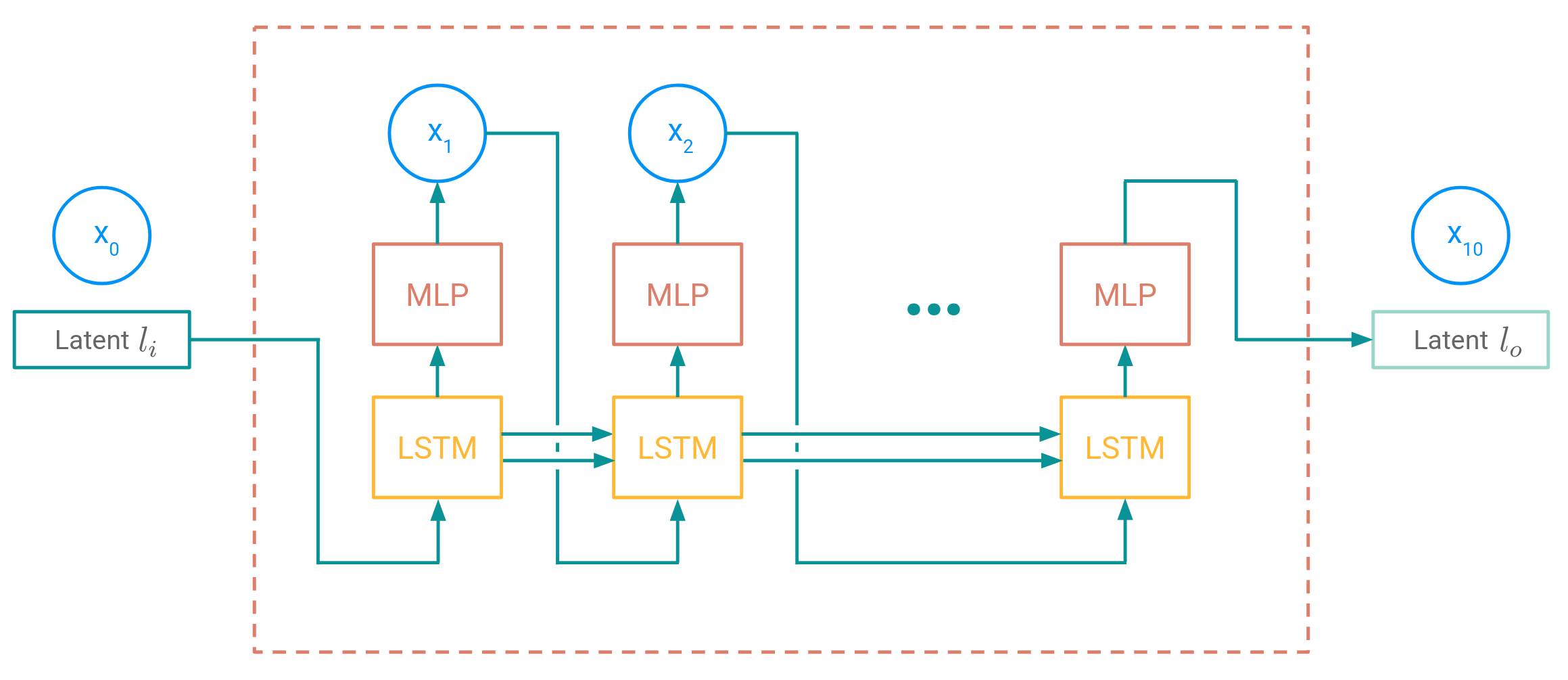}
    \caption{The design of the core network. It consists of a one-to-many LSTM + MLP configuration that is run for 10 steps.}
    \label{fig:core_lstm}
\end{figure*}

The design of the core network is visualised in Figure \ref{fig:core_lstm}. We interpret the latent representation of the inputs $l_i$ as the initial value $x_0$. The LSTM and MLP configuration produces 9 intermediary 'evolved' latent representations $\{x_1, ..., x_9\}$ before arriving at the final evolved latent representation $x_{10}$. We interpret this latent representation as the prediction of the latent representation of the evolved output $l_o$.

\subsubsection*{Training}

When the core model predicts a sequence of 10 latent representations, it is essentially traversing the latent space. We can guide the network to learn to traverse the latent space similarly to how VULCAN simulations evolve by using the sequence of outputs saved in the dataset (Section \ref{sec:dataset}). We do this in two ways: first, we construct a loss function that not only depends on the accuracy of the prediction of the latent representation of the final output $l_o$, but also on the accuracy of the intermittent latent representation predictions:

\begin{equation}
    \mathcal{L} = \sum_{t=1}^{10} \left(  \frac{1}{N} \sum_{i=1}^{N} \left ( p_{t,i} - a_{t,i} \right)^{2} \right),
\end{equation}

where $\mathcal{L}$ is the loss, $N$ is the number of elements in the actual/predicted vector, $p_{t,i}$ is the $i$-th element of the latent representation prediction vector at time step $t$, and $a_{t,i}$ is the $i$-th element of the latent representation vector of the output at time step $t$. With this notation, $a_{10} = l_o$. By training a network with this loss function, we force the core network to evolve the latent mixing ratios similarly to how VULCAN evolves mixing ratios. It should be noted that the latent representation of the inputs $l_i$ is larger than the latent representation of the outputs $l_o$, as it contains more properties than just mixing ratios. The predicted latent representations $x_t$ are therefore also larger than $l_o$. To account for this, we only look at the elements corresponding to the encoded mixing ratios in $l_i$ when comparing the predicted latent representations $x_t$ and the output mixing ratios $a_t / l_o$.

To further incentivise the core network to adhere to VULCAN's evolution patterns, we can intercept the predicted latent representations $x_t$ before they get fed back into the LSTM, and replace them with the latent representation of the actual output of the corresponding time step. This way, the core network is always learning from latent representations that follow VULCAN's evolution, even if the network is predicting poorly. This is only done during the training of the network when the true VULCAN outputs are known. 
During validation and testing, the predicted latent representations $x_t$ are not altered.

The core model LSTM has a hidden- and cell size of 4096. The MLP has only two layers: an input layer of size 4096, and an output layer of the same size as the latent representation of the inputs $l_i$, followed by a hyperbolic tangent function. It is optimised with the Adam optimiser, with a learning rate of $10^{-4}$ and a batch size of 8. It is trained for 100 epochs.

\subsubsection{Deployment}

When the trained model is deployed on the validation- and test dataset, the first step is encoding the inputs into their latent representation $l_i$ using the encoder part of the autoencoder (top left section in figure \ref{fig:model_overview}). The core network then predicts the latent representation of the evolved VULCAN output $l_o$ by traversing the latent space in 10 steps (centre section in figure \ref{fig:model_overview}). The prediction of the latent representation of the VULCAN output is then decoded by the decoder part of the autoencoder to obtain the predicted mixing ratios (bottom right section in figure \ref{fig:model_overview}).

\section{Results}
\label{sec:results}

\subsection{Autoencoders}

\subsubsection{Mixing Ratio Autoencoder}

\begin{figure*}
    \centering
    \includegraphics[scale=0.83]{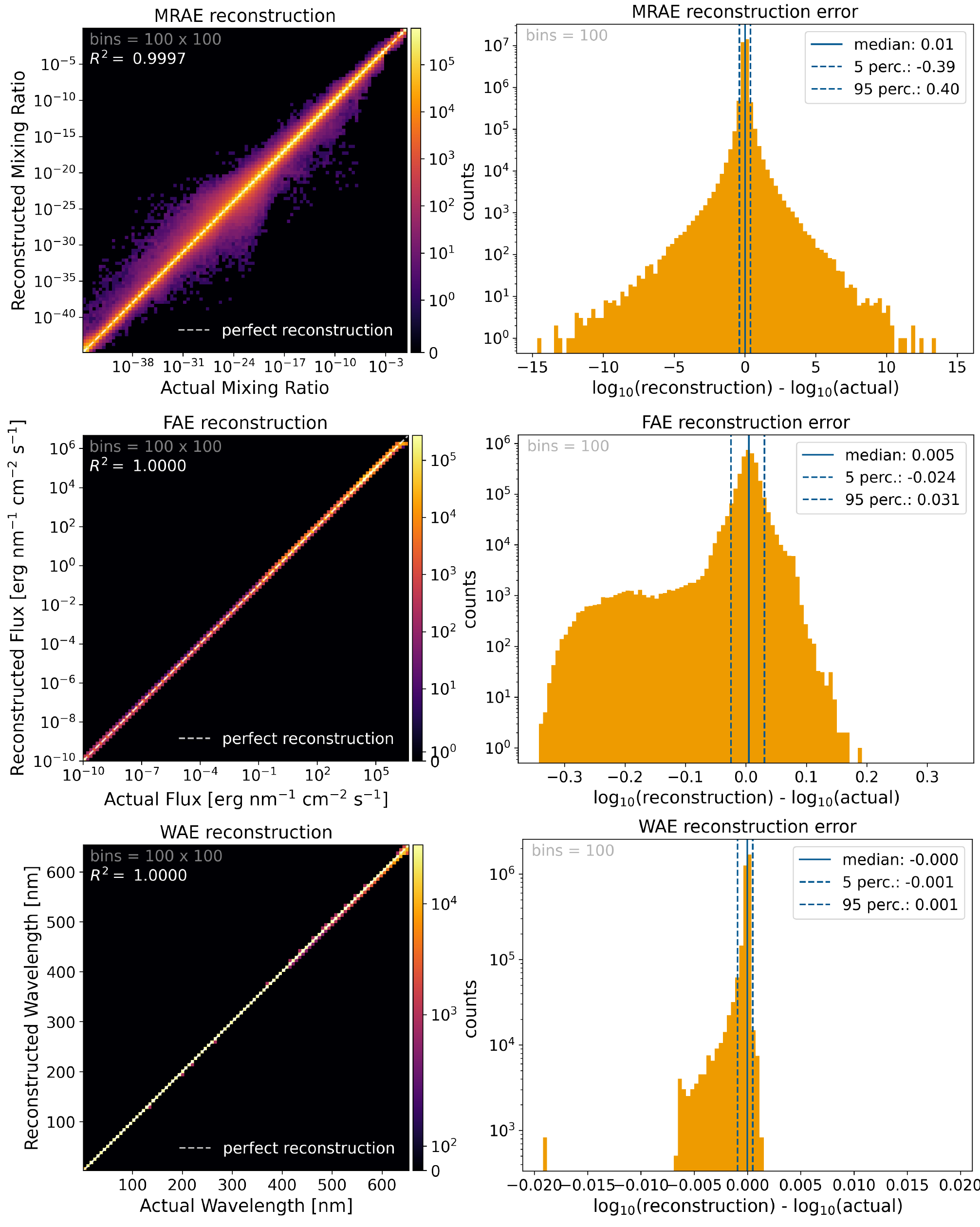}
    \caption{The reconstructed against the actual input values (left column), and the reconstruction error in log space (right column) for the mixing ratios (top row), stellar flux (middle row), and the wavelengths (bottom row). The diagonal dashed line in the reconstructed vs. actual mixing ratios plot shows the performance of a perfectly reconstructing model. Here the colour represents the number of examples within each bin. In the reconstruction error figure, the solid line shows the median value, and the dashed lines show the 5th and 95th percentiles. The R$^2$ values of each reconstruction plot are shown in the left column.}
    \label{fig:AE_part_A}
\end{figure*}

\begin{figure*}
    \centering
    \includegraphics[scale=0.83]{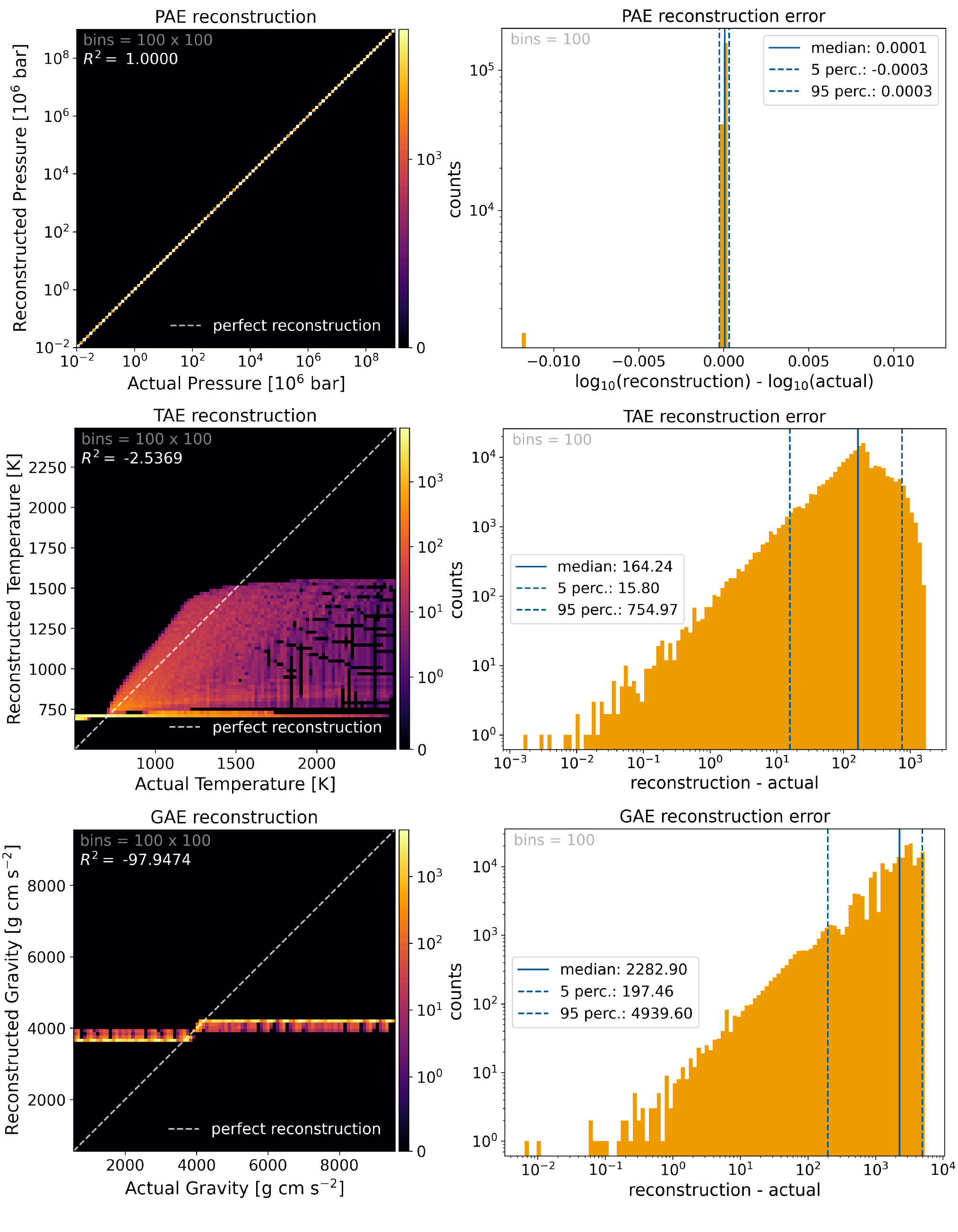}
    \caption{The reconstructed against the actual input values (left column), and the reconstruction error in log space (right column) for the pressure profile (top row), temperature profile (middle row), and the gravity profile (bottom row). Note that the reconstruction errors plot of the temperature and gravity values are calculated in linear space.}
    \label{fig:AE_part_B}
\end{figure*}

The top row of figure \ref{fig:AE_part_A} shows the reconstructed mixing ratio values against their actual value (left plot), for all examples from the test dataset. For the entire range of mixing ratios, the majority of the reconstructions lie within an order of magnitude of the diagonal line that marks perfect reconstructions, with an R-squared value of $R^2 = 0.9997$.

The right plot of the top row of figure \ref{fig:AE_part_A} shows the reconstruction error of the mixing ratios in logarithmic space. This scale is chosen because the autoencoders are trained in log space (see section \ref{sec:dataset}). The solid line shows the median and the dashed lines show the 5th and 95th percentiles. From the right figure, we can see that 90\% of the reconstructions have an error between -0.39 and 0.40 orders of magnitude.

\subsubsection{Flux Autoencoder}

The middle row of figure \ref{fig:AE_part_A} (left) shows the reconstructed flux values against their actual value. All reconstructed flux values are within 0.5 order magnitude of the graph diagonal. At fluxes with values around $10^6$ erg nm$^{-1}$ cm$^{-2}$ s$^{-1}$, the FAE is slightly underpredicting the actual flux values.

From the reconstruction error plot (right) we can see that 90\% of the reconstructions have an error between -0.024 and 0.031 orders of magnitude. In this figure, we see a distinct underprediction of a small number of examples, which are the high flux values we see being underpredicted.

\subsubsection{Wavelength Autoencoder}

The bottom row of figure \ref{fig:AE_part_A} (left) shows the reconstructed wavelength values against their actual value. All reconstructed wavelength values are close to the actual values, deviating less than $\sim 10$ nm from the graph diagonal. We can see that for wavelengths with values around $650$ nm, the FAE has a tendency to underpredict.

From the reconstruction error of the wavelength values plot (right), we can see that 90\% of the reconstructions have an error between -0.001 and 0.001 orders of magnitude. The slight underprediction of higher wavelength values is also visible in this figure.

\subsubsection{Pressure Profile Autoencoder}

The top row of figure \ref{fig:AE_part_B} (left) shows the reconstructed pressure values against their actual value. All reconstructed pressure values are well within 0.1 order magnitude of the graph diagonal. 

From the reconstruction error plot of the pressure values (right) we can see that 90\% of the reconstructions have an error between -0.0003 and 0.0003 orders of magnitude. This figure also shows a very minor underprediction of some pressure values, which correspond to pressure values of $\sim 10^3$ bar.

\subsubsection{Temperature Profile Autoencoder}

The middle row of figure \ref{fig:AE_part_B} (left) shows the reconstructed temperature values against their actual value, for samples from the test dataset. It is immediately obvious that this autoencoder cannot accurately reconstruct the temperature profiles. A fraction of temperatures in the range $\sim750$ K < T < $\sim1400$ K are reconstructed close to the graph diagonal, but the TAE is largely overpredicting temperatures below $\sim750$ K and underpredicting temperatures above $\sim750$ K. 

The histogram of reconstruction errors of the temperature values (right) shows under- and over-prediction. 90\% of the reconstructions have an error between 15.8 K and 754.97 K. Most predictions outside this range are underpredictions.

\subsubsection{Gravity Profile Autoencoder}

The GAE shows similar behaviour to the TAE. The bottom row of figure \ref{fig:AE_part_B} (left) shows the reconstructed gravity values against their actual value, for samples from the test dataset. Gravity values below $\sim 4000$ cm s$^{-2}$ are underpredicted by the autoencoder, while values above $\sim 4000$ cm s$^{-2}$ are underpredicted. Only values around $\sim 4000$ cm s$^{-2}$ are predicted accurately by the GAE. In the reconstruction errors plot of the GAE of figure \ref{fig:AE_part_B} (right) we can see that gravity values are consistently being over- and underpredicted. 90\% of the reconstructions have an error between 197.46 and 4939.6 cm s$^{-2}$. Most predictions outside this range are over-predictions.

\subsection{Core Network}

Because the TAE and GAE do not accurately reconstruct the temperature and gravity profiles, these profiles were not encoded for the final model. Instead, they were put directly in the latent representations of the inputs. This way, no information contained in these profiles is lost.

\begin{figure*}
    \centering
    \includegraphics[scale=0.85]{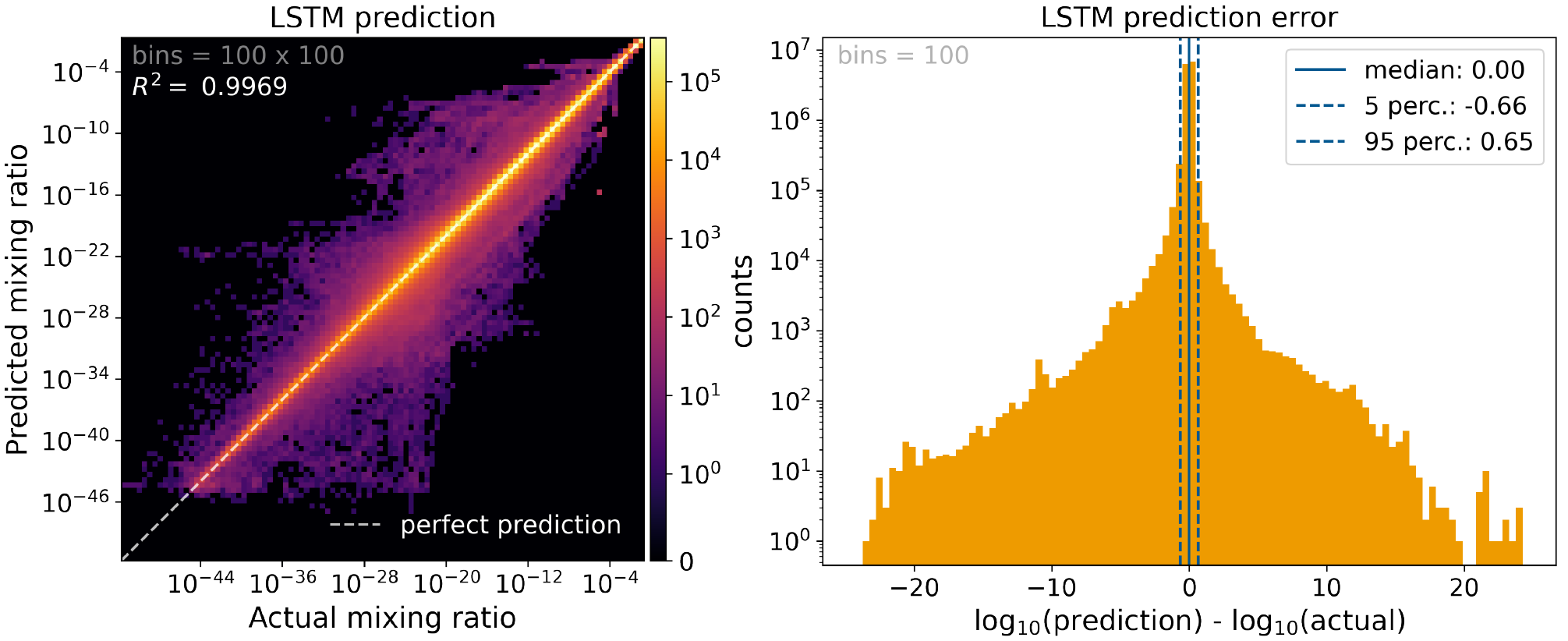}
    \caption{The LSTM predicted mixing ratios plotted against the actual mixing ratios (left) and the LSTM mixing ratio prediction error in log space (right). The dashed diagonal line in the left plot shows the performance of a perfectly predicting model and the colour of each bin represents the number of predictions. The solid line in the right plot shows the median value, and the dashed lines show the 5th and 95th percentiles.}
    \label{fig:LSTM_results}
\end{figure*}

The left plot in figure \ref{fig:LSTM_results} shows the mixing ratios predicted by the trained neural network model against the actual mixing ratios for the test dataset. The histogram shows that most of the model predictions lie within $\sim 1$ order magnitude of the diagonal of the graph. A notable exception is the predictions for the few mixing ratios with values lower than $\sim 10^{-44}$, for which the model overpredicts. These species can be neglected since they are not abundant enough to play a big role in the chemistry or to show features in the observable spectra.

\begin{figure*}
    \centering
    \includegraphics[scale=1.0]{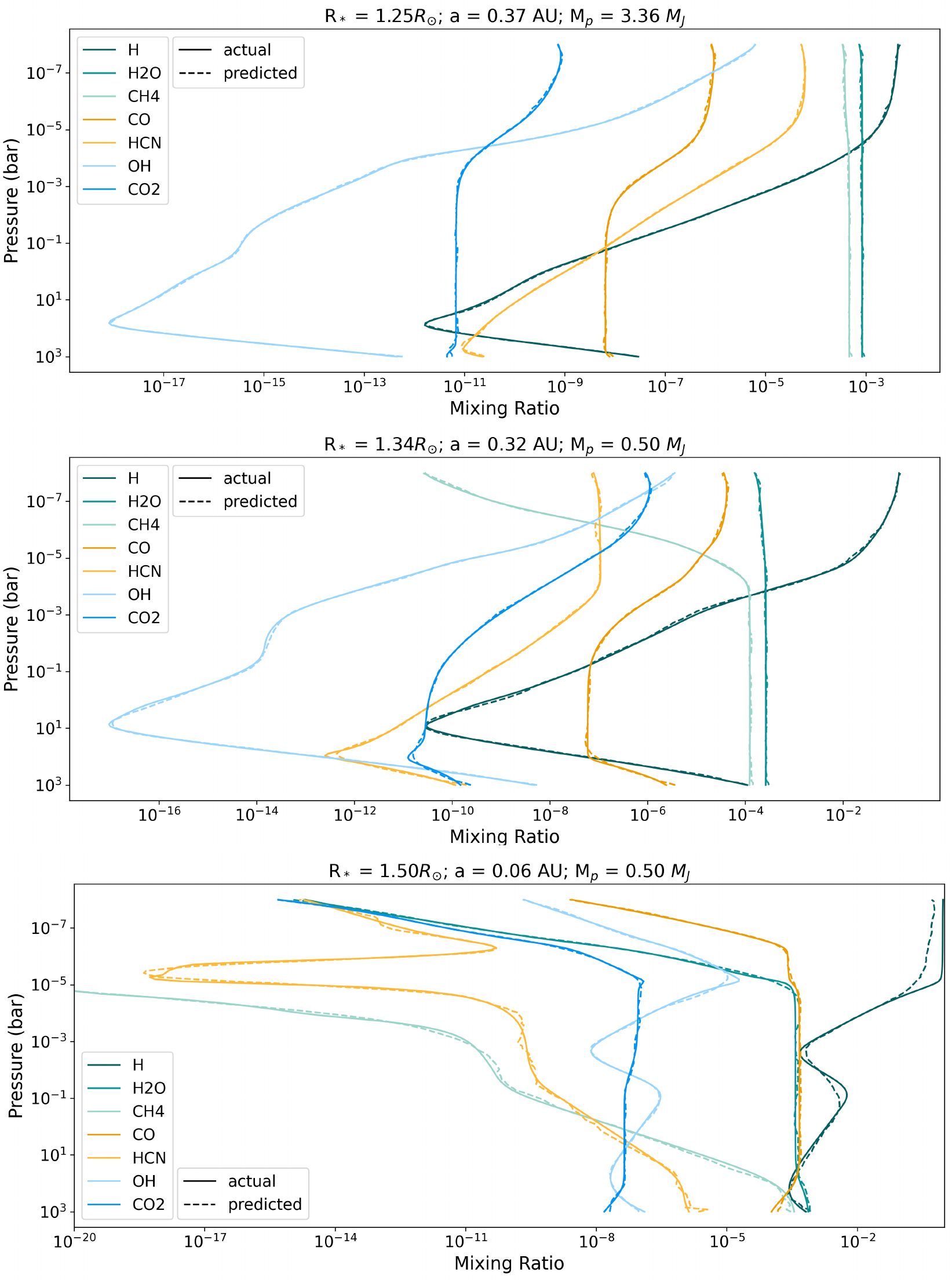}
    \caption{The mixing ratios per height layer for the best (top), typical (middle), and the worst (bottom) case of the validation set. The planet parameters for each case are given at the top of the plot. The solid lines show the actual mixing ratios as calculated by VULCAN, and the dashed lines show the neural network model predictions.}
    \label{fig:core_example}
\end{figure*}

Figure \ref{fig:LSTM_results} (right) shows the mixing ratio prediction error of the neural network model, in log-space. The solid line shows the median and the dashed lines show the 5th and 95th percentiles. From the figure, we can see that 90\% of the model predictions have an error between -0.66 and 0.65 orders magnitude. Outside of this range, the model does not show a clear tendency to either over- or underpredict.

Figure \ref{fig:core_example} shows selected examples (best, typical, and worst cases) of predictions by the neural network compared with the output of the VULCAN model, for a selection of seven species. The best case (top panel) shows a prediction that is almost indistinguishable from the actual mixing ratios. 
The examples in the typical case (middle panel) and worst case (lower panel) show larger prediction errors. In the typical case, CO$_2$, CO, and HCN have the largest prediction errors in the lower atmosphere, though still negligible. The worst case shows the largest prediction errors, with H having prediction errors of up to almost 1 order of magnitude in the upper atmosphere. Notable is that this case has very strong photochemistry in the upper atmosphere as it's positioned nearby its host star.

To compare the computational efficiency of VULCAN and the neural network model, the computational time to calculate or predict every example in the full dataset was recorded. The results are presented in table \ref{tab:computational_times}. It should be noted that the VULCAN simulations were run on similar, but older hardware than the neural network model. The median computational times show a $\sim 7.5\cdot 10^3\times$ decrease in computational time for the neural network model. The longest computational time required by the neural network model still shows a $\sim 10^3\times$ decrease in computational time compared to the fastest VULCAN simulation.

\begin{table}
    \centering
    \caption{Median, minimum and maximum running times of VULCAN and the neural network model for all configurations in the dataset. VULCAN runs were performed on a single CPU core using an \texttt{Intel(R) Xeon(R) CPU E5-4620 0 @ 2.20GHz}. The neural network model was run on a single core using an 
    \texttt{Intel(R) Xeon(R) W-1250 CPU @ 3.30GHz}}.
    \begin{tabular}{c|c|c|c}
         \textbf{code} & \textbf{median} & \textbf{minimum} & \textbf{maximum} \\
         \hline
         \hline
         VULCAN & 5994.3 s & 1236.7 s & 102223.0 s\\
         NN model & 0.77 s & 0.73 s & 0.93 s\\
    \end{tabular}
    \label{tab:computational_times}
\end{table}

\section{Discussion}
\label{sec:discussion}

In this study, we successfully used autoencoders to extract most of the characterising input features and encode them into latent representations for the mixing ratios, stellar flux, wavelengths, and pressure profiles.
Within these four groups, the largest prediction errors stem from the MRAE due to the high variability in input values, as opposed to the other input sources. We included initial and evolved mixing ratios of 69 species over 150 height layers. Additionally, the mixing ratio profiles among species differed significantly from one another (e.g. CH$_4$ and CO in figure \ref{fig:core_example}). This made the complexity, of extracting and encoding the fundamental input features, highest for this particular autoencoder. In opposition, the variety in the flux from stellar spectra was much less. We obtained the stellar spectra from either the MUSCLES database or generated them using the PHOENIX model. The spectra from these different sources were quite distinct from each other in the EUV (0.5 - 200 nm). The PHOENIX models assume the spectra to follow blackbodies, while, in reality, M- and K stars have shown to be highly active in the EUV (\cite{activity}), as was observed by the MUSCLES collaboration. Nonetheless, the spectra within each method seemed largely similar, which made it more straightforward for the FAE to learn how to accurately reconstruct them. For both the PAE and the WAE, the profiles were linearly spaced in logarithmic space, which made it easy for the autoencoders to learn how to encode these parameters. It is remarkable, however, that the WAE was not able to perfectly reproduce the wavelengths. A solution would be to make use of a handcrafted algorithm that encodes merely the first and last elements in the array. It is recommended to make use of such an algorithm for future use. Finally, the temperature- and gravity profile autoencoders were not successful at encoding and reconstructing their inputs. Both autoencoders produced the same solutions for each input example. The limited data set size and large variations in the temperature and gravity example cases could explain the autoencoders to be prone to errors. Future studies could focus on improving these specific autoencoders by performing root cause analysis.
However, a more specialised approach to encoding the pressure and gravity profiles would be to provide hyperparameters, such as the planet mass and radius, directly to the core model. Such an approach negates the need to train autoencoders for these input parameters.

\begin{figure}
    \centering
    \includegraphics[scale=0.5]{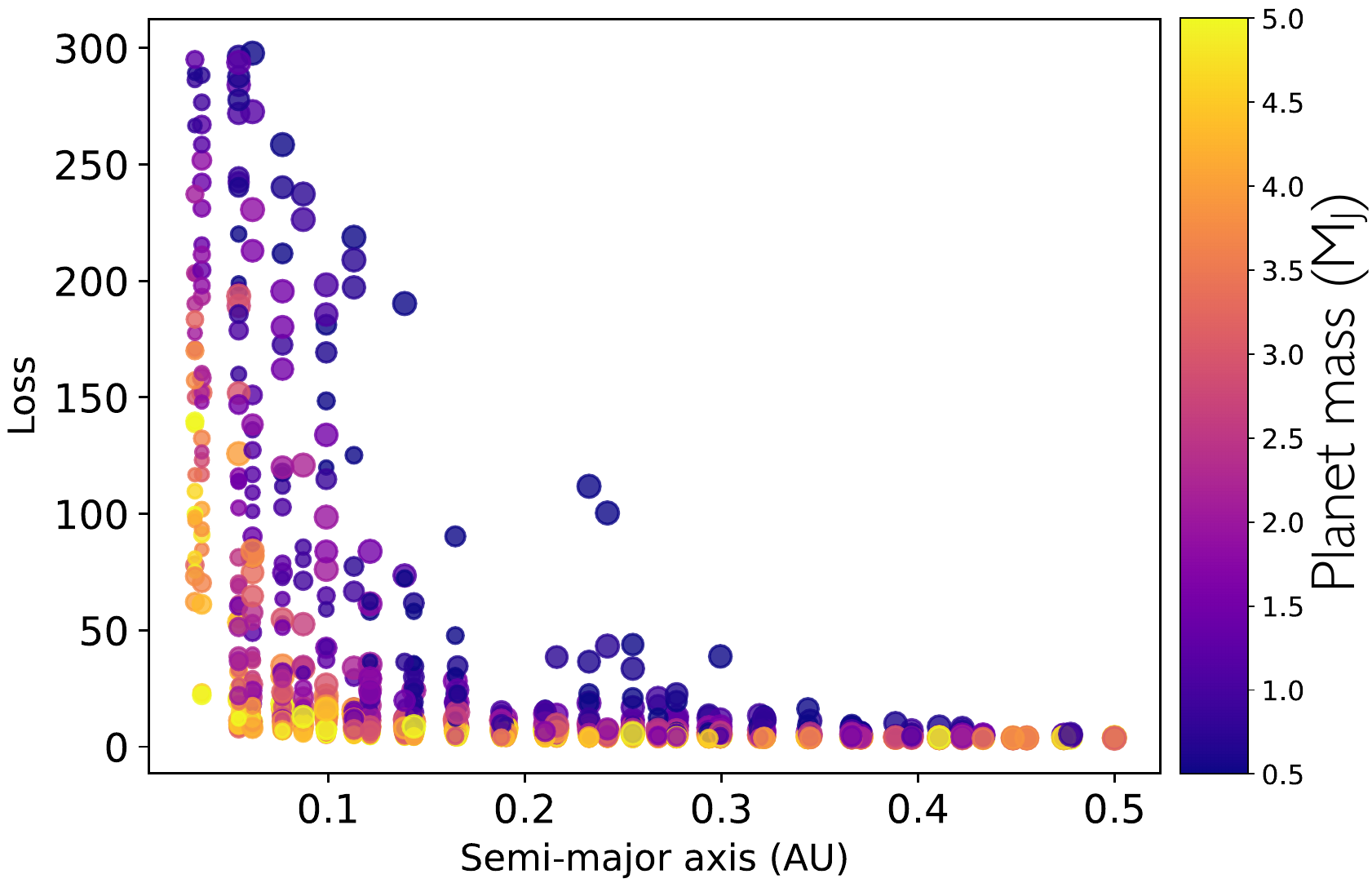}
    \caption{The loss as a function of the semi-major axis of each validation case. The colour represents the planet mass in $M_J$ and the size of each scatter point represents the size of the host star which ranges between 1 $M_{\odot}$ and 1.5 $M_{\odot}$. The loss is calculated by making use of eq. \ref{eq:relative MSE}}. 
    \label{fig:correlation_Loss_params}
\end{figure}

The prediction of the core network (LSTM) is within one order of magnitude for the majority (>90\%) of the predictions. These errors are comparable with the discrepancies between different chemical kinetics codes \citep{Venot12}. However, the accuracy of predictions of different examples varies. This inconsistency can arise due to some bias within the data set. Example cases similar to the best-case scenario (see \ref{fig:core_example}) were more prevalent in the data set, causing the core network to produce better predictions of this type of hot-Jupiters. Additionally, by plotting the loss of each validation case against input parameters (see figure \ref{fig:correlation_Loss_params}) it becomes apparent that some specific system parameters perform better than others. From figure \ref{fig:correlation_Loss_params}, we see that planets with smaller orbit radii seem to have worse predictions. One explanation could be that these planets endure more irradiation from their host star, ensuring photochemistry being the dominant process in the upper atmosphere. The abrupt and severe changes in abundances for some species due to photodissociation in the upper atmosphere could be difficult for the core network to learn with a limited dataset as provided in this study. Also noteworthy is the correlation between the planetary mass and the performance of the core network. Higher-mass planets tend to have lower losses as compared to lower-mass planets. Future work could focus on improving the prediction losses of the chemistry profiles for lower-mass planets and planets that orbit their host star close in.

We also showed that the trained model consistently over-predicts mixing ratios that have a value lower than $10^{-44}$. This can again be explained by the lack of examples that have such lower values. Species with mixing ratios this low are small contributors to the atmospheric composition nevertheless and are not expected to affect forward models.

Finally, we want to note that the hyperparameters used in this study have been found by trial and error and have not been proven to be the most optimal values. Future studies could focus on a hyperparameter search for each individual autoencoder and core network to find the most optimal parameters.

Due to all mentioned caveats, there is room for improvement in future work. Here we detail some of the aspects that are out of the scope of this paper, but we will be looking into them in future publications. Evidently, a larger size of the data set, with which the network is trained, is expected to improve the results significantly. In order to train a neural network to be more generalised and less biased, a more diverse and extensive data set should be created. Free parameters that can be taken into account, which were not explored in this study, are variables such as the eddy diffusion coefficient \footnote{Note that vertical mixing is taken into account in every simulation, but is kept constant throughout the dataset.}, condensation, and composition of the atmosphere (e.g. varying metallicity and the C/O ratio). 

Another approach to improve possibly the results is to change the model itself. The traditional autoencoders can be replaced with \textit{variational autoencoders}, VAEs \citep{VAE}. These types of autoencoders are based on Bayesian statistics. It is possible to regulate the latent space such that similar input examples have similar latent representations that lie close to each other within the latent space. The core network then might be able to learn how to traverse a regulated latent space and predict more accurately. The core network itself can be improved by, for example, including more time steps within the LSTM. Adding more time steps will ensure that the network predicts the solutions in a more similar way as VULCAN integrates toward the solution. A disadvantage of this is that the training time for the network will increase. Lastly, the recurrent neural network architecture could be changed to a \textit{transformer} design. Recently, the transformer neural network architecture, proposed by \cite{transformer}, revolutionised the field of sequence transactions within machine learning. By using a so-called \textit{attention} mechanism, the transformer neural network outperforms recurrent neural networks in accuracy and efficiency. Because of the similarity between transformer and recurrent neural network applications, the core model may perform better when changing to this new type of architecture. However, because transformers require tokenized inputs, the autoencoders will also have to be changed to produce the expected outputs. The implementation of a transformer would therefore increase the complexity of the entire model and should be done carefully.

A different direction in approach would be, for example, to use interpolation methods in the already existing data set. The limitation of such methods is the ease of distribution. The size of the data set used in this study is 600GB, 
as opposed to a size of 3GB for the weights of the neural network used in this study.

\section{Summary \& Conclusions}
\label{sec:conclusion}
In this study, we investigated the ability of a neural network to replace the time-dependent ordinary differential equations in the chemical kinetics code VULCAN (\citeauthor{Tsai17} \citeyear{Tsai17}; \citeyear{Tsai21}). The aim of this research was to explore the LSTM architecture for solving ordinary differential equations that include vertical mixing and photo-chemistry. 
We first created a data set that contains the in- and outputs of VULCAN simulations of hot-Jupiter atmospheres. We made use of the planetary mass (0.5 - 20 $M_{\mathrm{J}}$), the semi-major axis (0.01 - 0.5 AU), and the stellar radius (1 - 1.5 $R_{\odot}$) as free parameters. Other parameters for the VULCAN configurations were derived either from analytical relations
or kept constant throughout the data set. The input of the data set comprises the initial mixing ratios, the stellar spectrum, the temperature- and pressure profiles, and the gravity profiles. Note that the neural network trained in this study is limited to the chosen free parameters and can not be used for atmospheric models that include e.g. condensation. The outputs of the data set contain the mixing ratios of the species in the atmosphere, taken from 10-time steps (including the steady state) during the VULCAN simulation. This data was used to train a neural network that consists of two parts: the \textit{autoencoder} network and the \textit{core} network. The autoencoder was used to reduce the dimensionality of the input and output data from the data set by encoding them into lower dimensionality \textit{latent representations}. The autoencoder network consisted of six smaller autoencoders, designed and trained to encode and decode the mixing ratios, flux, wavelengths, and temperature-, pressure-, and gravity profiles to and from their respective latent representations. The total input latent representation was the concatenation of these 6 smaller ones. The core network was designed to have an LSTM-based architecture and it mapped between the latent representation of the inputs to the encoded evolved output by traversing the \textit{latent space} in ten steps. During the training, the latent representations at these ten steps were compared to the ten sets of mixing ratios saved in the outputs of the data set to ensure that the core network is evolving the latent representation in a similar fashion as the VULCAN simulation evolves the mixing ratios. To summarise, we found that:
\begin{itemize}
    \item the mixing ratios, flux, wavelengths, and pressure profile autoencoders were able to efficiently encode and accurately reconstruct their respective input properties
    \item the autoencoders were not able to encode and decode the temperature- and gravity profiles successfully. These autoencoders were, therefore, not used and instead, these profiles were put directly into the latent representation of the inputs
    \item the fully trained model (i.e. including the core network) was able to predict the mixing ratios of the species with the errors in the range [-0.66, 0.65] orders magnitude for 90\% of the cases. Due to imbalances in the dataset, the model is biased to more accurately solve for some examples as compared to others
    \item the fully trained model is $\sim 10^3$ times faster than the VULCAN simulations
\end{itemize}
Overall, this study has shown that machine learning is a suitable approach to accelerate chemical kinetics codes for modelling exoplanet atmospheres.

\section*{Data Availability}

All simulated data created in this study will be shared upon reasonable request to the corresponding author. The code and results are publicly available on \href{https://github.com/JuliusHendrix/MRP/tree/main}{github.com/JuliusHendrix/MRP}.



\bibliographystyle{mnras}
\bibliography{main}




\appendix




\bsp	
\label{lastpage}
\end{document}